\documentclass[twocolumn,showpacs,aps]{revtex4}
\usepackage{amsmath}
\usepackage{graphicx}

\begin{document}
\title{Kinetic Theory of Random Graphs: from Paths to Cycles}

\author{E.~Ben-Naim}
\affiliation{Theoretical Division and Center for Nonlinear
Studies, Los Alamos National Laboratory, Los Alamos, New Mexico
87545}
\author{P.~L.~Krapivsky}
\affiliation{Center for Polymer Studies and Department of Physics,
Boston University, Boston, Massachusetts 02215}

\begin{abstract}

  Structural properties of evolving random graphs are investigated.
  Treating linking as a dynamic aggregation process, rate equations
  for the distribution of node to node distances (paths) and of cycles
  are formulated and solved analytically.  At the gelation point, the
  typical length of paths and cycles, $l$, scales with the component
  size $k$ as $l\sim k^{1/2}$.  Dynamic and finite-size scaling laws
  for the behavior at and near the gelation point are
  obtained. Finite-size scaling laws are verified using numerical
  simulations.

\end{abstract}
\maketitle

\section{Introduction}

A random graph is a set of nodes that are randomly joined by links. When
there are sufficiently many links, a connected component containing a finite
fraction of all nodes, the so-called giant component, emerges. Random graphs,
with varying flavors, arise naturally in statistical physics, chemical
physics, combinatorics, probability theory, and computer science
\cite{sr,er,bb,jlr,jklp}.

Several physical processes and algorithmic problems are essentially
equivalent to random graphs. In gelation, monomers form polymers via
chemical bonds until a giant polymer network, a ``gel'', emerges.
Identifying monomers with nodes and chemical bonds with links shows
that gelation is equivalent to the emergence a giant component
\cite{pjf,whs,pjf1}. A random graph is also the most natural
mean-field model of percolation \cite{ds,kcbh}.  In computer science,
satisfiability, in its simplest form, maps onto a random graph
\cite{bbckw}. Additionally, random graphs are used to model social
networks \cite{nsw,gn}.

Random graphs have been analyzed largely using combinatorial and
probabilistic methods \cite{bb,jlr,jklp}. An alternative
statistical physics methodology is kinetic theory, or
equivalently, the rate equation approach. The formation of
connected components from disconnected nodes can be treated as a
dynamic aggregation process \cite{mvs,chandra,da,fl}. This kinetic
approach was used to derive primarily the size distribution of
components \cite{jbm,hez,bk}.

Recently, we have shown that structural characteristics of random
graphs can be analyzed using the rate equation approach \cite{bk1}. In
this study, we present a comprehensive treatment of paths and cycles
in evolving random graphs. The rate equation approach is formulated by
treating linking as a dynamic aggregation process. This approach
allows an analytic calculation of the path length distribution.  Since
a cycle is formed when two connected nodes are linked, the path length
distribution yields the cycle length distribution. More subtle
statistical properties of cycles in random graphs can be calculated as
well. In particular, the probability that the system contains no
cycles and the size distribution of the first, second, etc. cycles are
obtained analytically.

We focus on the behavior near and at the phase transition point,
namely, when the gel forms. We show that the path and the cycle length
distribution approach self-similar distributions near the gelation
transition. At the gelation point, these distributions develop
algebraic tails.

The exact results obtained for an infinite system allow us to deduce
scaling laws for finite systems. Using heuristic and extreme
statistics arguments, the size of the giant component at the gelation
point is obtained. This size scale characterizes the size distribution
of components and it leads to a number of scaling laws for the typical
path size and cycle size. Extensive numerical simulations validate
these scaling laws for finite systems.

The rest of the paper is organized as follows. First, the evolving
random graph process is introduced (Sec.~II), and then the size
distribution of all components is analyzed in Sec.~III.
Statistical properties of paths are derived in Sec.~IV and then
used to obtain statistical properties of all cycles (Sec.~V) and
of the first cycle (Sec.~VI). We conclude in Sec.~VII. Finally, in
an appendix, some details of contour integration used in the body
of the paper are presented.

\section{Evolving Random graphs}

A graph is a collection of nodes joined by links. In a random graph,
links are placed randomly. Random graphs may be realized in a number
of ways.  The links may be generated instantaneously (static graph) or
sequentially (evolving graph); additionally a given pair of nodes may
be connected by at most a single link (simple graph) or by multiple
links (multi-graph).

\begin{figure}[ht]
\includegraphics[width=4cm]{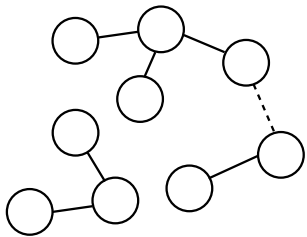}
\caption{An evolving random graph. Links are indicated by solid lines
and the newly added link by a dashed line.}
\label{fig-rg}
\end{figure}

We consider the following version of the random graph model.
Initially, there are $N$ disconnected nodes. Then, a pair of nodes is
selected at random and a link is placed between them
(Fig.~\ref{fig-rg}). This linking process continues ad infinitum and
it creates an evolving random graph. The process is realized
dynamically. Links are generated with a constant rate in time, set
equal to $(2N)^{-1}$ without loss of generality. There are no
restrictions associated with the identity of the two nodes.  A pair of
nodes may be selected multiple times, i.e., a multi-graph is
created. Additionally, the two nodes need not be different, so
self-connections are allowed.

At time $t$, the total number of links is on average $Nt/2$, the
average number of links per node (the degree) is $t$, and the average
number of self-connections per node is $N^{-1}t/2$. Therefore, whether
or not self-connections are allowed is a secondary issue.  Since the
linking process is completely random, the degree distribution is
Poissonian with a mean equal to $t$.

\section{Components}

The evolving random graph model has several virtues that simplify the
analysis.  First, the linking process is completely random as there is
no memory of previous links. Second, having at hand a continuous
variable (time) allows us to use continuum methods, particularly the
rate equation approach. This is best demonstrated by determination of
the size distribution of connected components.

As linking proceeds, connected components form. When a link is placed
between two distinct components, the two components join.  For
example, the latest link in Fig.~\ref{fig-rg} joins two components of
size $i=2$ and $j=4$ into a component of size $k=i+j=6$. Generally,
there are $i\times j$ ways to join disconnected components. Hence,
components undergo the following aggregation process
\begin{equation}
\label{agg}
(i,j)\buildrel ij/2N\over \longrightarrow i+j.
\end{equation}
Two components aggregate with a rate proportional to the product of
their sizes.

\subsection{Infinite Random Graph}
\label{sub:inf}

Let $c_k(t)$ be the density of components containing $k$ nodes at time
$t$. In terms of $N_k(t)$, the total number of components with $k$
nodes, then $c_k(t)=N_k(t)/N$. For finite random graphs, both $N_k(t)$
and $c_k(t)$ are random variables, but in the $N\to\infty$ limit the
density $c_k(t)$ becomes a deterministic quantity.  It evolves
according to the {\it nonlinear} rate equation (the explicit time
dependence is dropped for simplicity)
\begin{equation}
\label{ck-eq}
\frac{dc_k}{dt}=\frac{1}{2}\sum_{i+j=k}(ic_i)(jc_j)-k\,c_k.
\end{equation}
The initial condition is $c_k(0)=\delta_{k,1}$.  The gain term
accounts for components generated by joining two smaller components
whose sizes sum up to $k$. The second term on the right-hand side of
Eq.~(\ref{ck-eq}) represents loss due to linking of components of size
$k$ to other components. The corresponding gain and loss rates follow
from the aggregation rule (\ref{agg}).

The rate equations can be solved using a number of
techniques. Throughout this investigation, we use a convenient method
in which the time dependence is eliminated first. Solving the rate
equations recursively yields $c_1=e^{-t}$, $c_2=\frac{1}{2}te^{-2t}$,
$c_3=\frac{1}{2}t^2e^{-3t}$, etc.  These explicit results suggest that
$c_k(t)=C_k\, t^{k-1}\,e^{-kt}$. Substituting this form into
(\ref{ck-eq}), we find that the coefficients $C_k$ satisfy the
recursion relation
\begin{equation}
\label{Ck-eq}
(k-1)\,C_k=\frac{1}{2}\sum_{i+j=k}(iC_i)\,(jC_j)
\end{equation}
subject to $C_1=1$. This recursion is solved using the generating
function approach. The form of the right-hand side of
Eq.~(\ref{Ck-eq}) suggests to utilize the generating function of the
sequence $kC_k$ rather than $C_k$, i.e., \hbox{$G(z)=\sum_k
k\,C_k\,e^{kz}$}.  Multiplying Eq.~(\ref{Ck-eq}) by $k\,e^{kz}$ and
summing over all $k$, we find that the generating function satisfies
the nonlinear ordinary differential equation
\begin{equation}
\label{g-eq}
(1-G)\,\frac{dG}{dz}=G.
\end{equation}
Integrating this equation, $z=\ln G-G+A$ and using the asymptotics
$G\to e^z$ as $z\to-\infty$ fixes the constant $A=0$. Thus, we arrive
at an implicit solution for the generating function
\begin{equation}
\label{gz}
G\,e^{-G}=e^z.
\end{equation}
The coefficients $C_k$ can be extracted from (\ref{gz}) via the
Lagrange inversion formula, or using contour integration as detailed
in Appendix A. Substituting $r=1$ in Eq.~(\ref{ak}) yields
$C_k=\frac{k^{k-2}}{k!}$ reproducing the well-known result for the
size distribution \cite{jbm,hez}
\begin{equation}
\label{ckt}
c_k(t)=\frac{k^{k-2}}{k!}\,t^{k-1}\,e^{-kt}.
\end{equation}

In the following, we shall often use the generating function for the
size distribution $c(z,t)=\sum_k k\, c_k(t) e^{kz}$. This generating
function is readily expressed via the auxiliary generating function 
$G(z)=\sum_k k\,C_k\,e^{kz}$:
\begin{equation}
\label{czt}
c(z,t)=t^{-1}G(z+\ln t-t).
\end{equation}

Let us consider the fraction of nodes in finite components,
$M_1=\sum_k k\,c_k(t)$.  This quantity is merely the first moment of
the size distribution (hence the notation). Equivalently
$M_1=c(z=0,t)$. {}From (\ref{czt}) we find $M_1=\tau/t$ with
$\tau=G(\ln t-t)$. Using (\ref{gz}), we express $\tau$ through $t$:
\begin{equation}
\label{tau}
\tau e^{-\tau}=te^{-t}.
\end{equation}
For $t<1$, there is a single root $\tau=t$, and all nodes reside in
finite components, $M_1=1$. For $t>1$ the physical root satisfies
$\tau<t$ and only a fraction of the nodes resides in finite
components, $M_1<1$. Thus, at time $t=1$, the system undergoes a
gelation transition with a finite fraction of the nodes contained in
infinite components. We term this time the gelation time, $t_g=1$. In
the late stages of the evolution $t\gg 1$, one has $\tau\simeq
te^{-t}$ and $M_1\simeq c_1= e^{-t}$, so the system consists of a
single giant component and a small number of isolated nodes.

The behavior at and near the transition point are of special
interest. The critical behavior of the component size distribution is
echoed by other quantities as will be shown below.  Size distributions
become algebraic near the critical point. Moreover, there is a
self-similar behavior as a function of time (dynamical scaling) and as
a function of the system size (finite-size scaling).

At the gelation point, the component size distribution has an
algebraic large-size tail, obtained using the Stirling formula,
\begin{equation}
\label{ckg}
{\bf c}_k\simeq C\,k^{-5/2}.
\end{equation}
with $C=(2\pi)^{-1/2}$.  [Throughout this paper, bold letters are used
for critical distributions, so ${\bf c}_k\equiv c_k(t=1)$.]  In the
vicinity of the gelation time, the size distribution is self-similar,
$c_k(t)\to (1-t)^5\Phi_c\big(k(1-t)^2\big)$ with the scaling function
\begin{equation}
\label{phic}
\Phi_c(\xi)=(2\pi)^{-1/2}\xi^{-5/2}\exp(-\xi/2).
\end{equation}
Thus, the characteristic component size diverges near the gelation point, 
$k\sim (1-t)^{-2}$.

\subsection{Finite Random Graphs}
\label{sub:finite}

In the previous subsection, we applied kinetic theory to an infinite
system. This approach can be extended to finite systems.
Unfortunately, such treatments are very cumbersome
\cite{aal,ve}. Since the number of components is finite, the
fluctuations are no longer negligible, and instead of a deterministic
rate equation approach, a stochastic approach is needed. Here we
follow an alternative path, employing the exact infinite system
results in conjunction with scaling and extreme statistics arguments.

The characteristic size of components at the gelation point exhibits
nontrivial dependence on the system size. This is conveniently seen
via the cumulative size distribution. The size of the largest
component in the system, $k_g$, is estimated from the extreme
statistics criterion, $N\sum_{k\geq k_g}{\bf c}_k\sim 1$, to be
\begin{equation}
\label{kg}
k_g\sim N^{2/3}.
\end{equation}
The largest component in the system grows sub-linearly with the system
size \cite{bb}. The time by which this component emerges approaches
unity for large enough systems as follows from the diverging characteristic
size scale $k_g\sim (1-t_g)^{-2}$,
\begin{equation}
\label{tg}
1-t_g\sim N^{-1/3}.
\end{equation}

\begin{figure}[ht]
\includegraphics*[width=8cm]{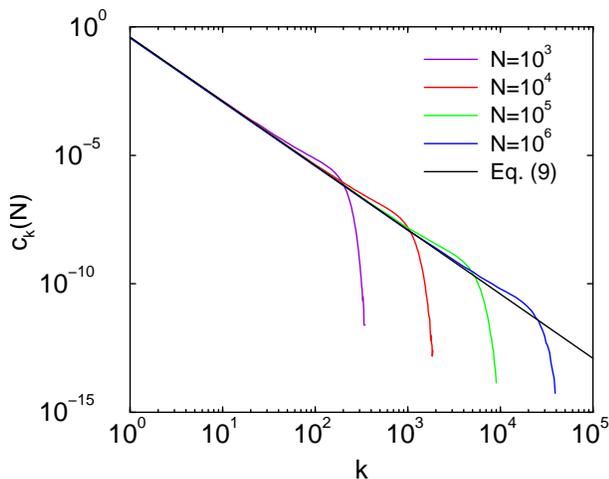}
\caption{The size distribution for a finite system at the gelation
  point. Shown is ${\bf c}_k(N)$ versus $k$ for various $N$. The infinite
  system behavior is shown for reference. The data represents an average over
  $10^6$ independent realizations.}
\label{fig-ckn}
\end{figure}

\begin{figure}[ht]
\includegraphics*[width=8cm]{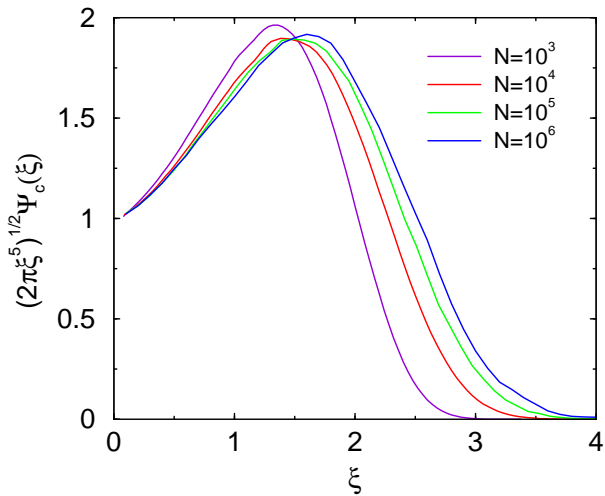}
\caption{Finite-size scaling of the size distribution. Shown is $(2\pi
  \xi^5)^{1/2}\Psi_c(\xi)$ versus $\xi$, obtained from simulations with
  various $N$.}
\label{fig-fss}
\end{figure}

The maximal component size (\ref{kg}) underlies the entire size
distribution. Let $c_k(N,t)$ be the size distribution in a system of
size $N$ at time $t$. At the gelation point, the size distribution
${\bf c}_k(N)\equiv c_k(N,t=1)$ obeys the finite-size scaling form
(Figs.~\ref{fig-ckn} and \ref{fig-fss})
\begin{equation}
\label{ckn}
{\bf c}_k(N)\sim N^{-5/3}\Psi_c(kN^{-2/3}).
\end{equation}
The scaling function has the following extremal behaviors
\begin{equation}
\label{psic}
\Psi_c(\xi)\simeq
\begin{cases}
(2\pi)^{-1/2}\xi^{-5/2}&\xi\ll 1;\\
\exp(-\xi^\gamma)&\xi\gg 1.
\end{cases}
\end{equation}

The small-$\xi$ behavior corresponds to sizes well below the characteristic
size and thus reflects the infinite system behavior (\ref{ckg}). The
large-$\xi$ behavior was obtained numerically with $\gamma\cong 3$. To
appreciate the large-$\xi$ asymptotic, let us estimate the probability
that the system managed to generate the largest possible component of
size $N/2$ at time $t=1$. The lower bound for this probability can be
established via a ``greedy'' evolution which assumes that after $k$
linking events the graph is composed of a tree of size $k+1$ and
$N-k-1$ disconnected nodes.  Such evolution occurs with probability
\begin{equation*}
\frac{2}{N}\cdot\frac{N-2}{N}\times
\frac{3}{N}\cdot\frac{N-3}{N}\times\ldots
\times\frac{N-N/2}{N}\cdot\frac{N/2}{N}\sim\frac{N!}{N^N},
\end{equation*}
that scales as $e^{-N}$. While this lower bound is not necessarily
optimal, it suggests that the actual probability is exponentially
small. The scaling variable $\xi=kN^{-2/3}$ becomes $\xi\sim N^{1/3}$
for $k=N/2$, so $\exp(-N^{\gamma/3})$ matches the probability
$\exp(-N)$ when $\gamma=3$.

To check the critical behavior in finite systems, we performed
numerical simulations.  In the simulations, $N/2$ links are placed
randomly and sequentially among the $N$ nodes as follows. A node is
drawn randomly, and then another node is drawn randomly. Last, these
two nodes are linked. Self-connections are therefore allowed. The
simulations differ slightly from the above random graph model in that
the number of links is not a stochastic variable. For large $N$, this
simulation is faithful to the evolving random graph model because the
number of links is self-averaging.

The simulation results are consistent with the postulated finite-size
scaling form (\ref{ckn}). We note that the scaling function
$\Psi_c(\xi)$ converges slowly as a function of $N$. The simulations
reveal an interesting behavior of the finite-size scaling
function. The function ${\bf c}_k(N)$ has a ``shoulder'' --- a
non-monotonic behavior compared with the pure algebraic behavior
(\ref{ckg}) characterizing infinite systems (Fig.~\ref{fig-ckn}). The
properly normalized scaling function $(2\pi \xi^5)^{1/2}\Psi_c(\xi)$
is a non-monotonic function of $\xi$ (Fig.~\ref{fig-fss}). Obtaining
the full functional form of the scaling function $\Psi_c(\xi)$ remains
a challenge.  A very similar shoulder has been observed for the degree
distribution of finite random networks generated by preferential
attachment \cite{zm,dms,bck,kr}.

\section{Paths}

Structural characteristics of components can be investigated in a
similar fashion. By definition, every two nodes in a component are
connected. In other words, there is a {\it path} consisting of
adjacent links between two such nodes. We investigate statistical
properties of paths in components. Characterization of paths
yields useful information regarding the connectivity of components
as well as internal structures such as cycles.

For every node in the graph, there are (generally) multiple paths that
connect it with all other nodes in the respective component.  With new
links, new paths are formed.  For every pair of paths of lengths $n$
and $m$ originating at two separate nodes, a new path is formed as
follows
\begin{equation}
\label{path}
n,m \longrightarrow n+m+1.
\end{equation}
In Fig.~\ref{fig-rg}, linking two paths of respective lengths $n=1$
and $m=2$ generates a path of length $n+m+1=4$. Thus, paths also
undergo an aggregation process. However, this aggregation process is
simpler than (\ref{agg}) because the aggregation rate is independent
of the path length.

Let $q_l(t)$ be the density of {\it distinct} paths containing $l$
links at time $t$. By distinct we mean that the two paths connecting
two nodes are counted separately. By definition, $q_0(t)=1$. The rest
of the densities grow according to the rate equation 
\begin{equation}
\label{ql-eq}
\frac{dq_l}{dt}=\sum_{n+m=l-1}q_nq_m
\end{equation}
for $l>0$.  The initial condition is $q_l(0)=\delta_{l,0}$. This rate
equation reflects the uniform aggregation rate. Another notable
feature is the lack of a loss term --- once a path is created, it
remains forever.  Solving recursively gives $q_1=t$, $q_2=t^2$,
etc. By induction, the path length density is
\begin{equation}
\label{ql}
q_l(t)=t^l.
\end{equation}
Indeed, this expression satisfies both the rate equation and the
initial condition. The first quantity $q_1=t$ is consistent with the
facts that the link density is equal to $t/2$ and that every link
corresponds to two distinct paths of length one.

The above path density represents an aggregate over all nodes and all
components. Characterization of path statistics in a component of a
given size is achieved via $p_{l,k}$, the density of paths of length
$l$ in components of size $k$.  Note the obvious length bounds $0\leq
l\leq k-1$ and the sum rule $\sum_l p_{l,k}=k^2 c_k$ reflecting that
there are $k^2$ distinct paths in a component of size $k$ (every pair
of nodes is connected). The density of the linkless paths is
$p_{0,k}=kc_k$, because $kc_k$ is the probability that a node belongs
to a component of size $k$.

We have seen that components and paths form via the aggregation
processes (\ref{agg}) and (\ref{path}), respectively. The joint
distribution $p_{l,k}$ therefore undergoes a bi-aggregation process
\cite{kb}. In the present case,
\begin{equation}
\label{bi-agg}
(n,i)+(m,j)\longrightarrow (n+m+1,i+j)
\end{equation}
where the first index corresponds to the path length and the second to the
component size. The joint distribution evolves according to the rate equation
\begin{equation}
\label{plk-eq}
\frac{dp_{l,k}}{dt}=\!\!\!\sum_{\substack{i+j=k\\n+m=l-1}}\!\!\!p_{n,i}p_{m,j}
+\sum_{i+j=k}(ip_{l,i})(jc_j)-kp_{l,k}.
\end{equation}
The initial conditions are $p_{l,k}(0)=\delta_{k,1}\delta_{l,0}$. The
first term on the right-hand side of Eq.~(\ref{plk-eq}) describes
newly formed paths due to linking. The last two terms correspond to
paths that do not contain the newly placed link.

We now repeat the steps used to determine the size distribution.  The
time dependence is eliminated using the ansatz
$p_{l,k}=P_{l,k}t^{k-1}e^{-kt}$.  The corresponding coefficients
$P_{l,k}$ satisfy the recursion
\begin{eqnarray}
\label{Plk-eq}
(k-1)P_{l,k}=\!\!\!\sum_{\substack{i+j=k\\n+m=l-1}}\!\!\!P_{n,i}P_{m,j}
+\sum_{i+j=k}(iP_{l,i})(jC_j).
\end{eqnarray}
The generating function $P_l(z)=\sum_k P_{l,k}e^{kz}$ satisfies the
recursion relation
\hbox{$(1-G)\,\frac{dP_l}{dz}=\sum_{n+m=l-1}P_{n}P_{m}+P_l$} for
$l>0$.  Dividing this equation by (\ref{g-eq}) yields
\begin{equation}
\label{Pl-eq}
G\,\frac{dP_l}{dG}=\sum_{n+m=l-1}P_{n}P_{m}+P_l
\end{equation}
for $l>0$. As noted above $P_{0,k}=kC_k$, so $P_0(z)=G(z)$.  Solving
Eq.~(\ref{Pl-eq}) recursively gives $P_1=G^2$, $P_2=G^3$, etc. In
general,
\begin{equation}
\label{Pl} P_l(z)=G^{l+1}(z).
\end{equation}
This solution can be validated directly. The time dependent generating
function $p_l(z)=\sum_k p_{l,k}e^{kz}$ is therefore
$p_l(z)=t^{-1}G^{l+1}(z+\ln t-t)$. The total density of paths of
length $l$, $p_l(z=0)=t^l$, coincides with (\ref{ql}) prior to the
gelation transition ($t<1$) because all components are finite.
However, the total number of paths is reduced,
$p_l(z=0)=t^{-1}\tau^{l+1}$, past the gelation time ($t>1$).

One may also obtain the bivariate generating function $p(z,w)=\sum_{l,k}
p_{l,k} w^l e^{kz}$. Using (\ref{Pl}) one gets
\begin{equation}
\label{pzw}
p(z,w)=t^{-1}\frac{G(z+\ln t-t)}{1-wG(z+\ln  t-t)}.
\end{equation}
The total density of paths in finite components is of course
$g=\sum_{l,k}p_{l,k}$, so $g\equiv p(z=0,w=1)$. Generally,
$g=\frac{\tau}{t(1-\tau)}$; for $t<1$ the total density of paths is
$g(t)=(1-t)^{-1}$.

The coefficients are found via the contour integration $P_{l,k}=(2\pi
i)^{-1}\oint dy\, P_l\,y^{-k-1}$ (see Appendix A). Substituting
$r=l+1$ in Eq.~(\ref{ak}) yields
$P_{l,k}=(l+1)\,\frac{k^{k-l-2}}{(k-l-1)!}$. As a result, the density
of paths of length $l$ in components of size $k$ is
\begin{equation}
\label{plk}
p_{l,k}=(l+1)\,\frac{k^{k-l-2}}{(k-l-1)!}\,t^{k-1}\,e^{-kt}.
\end{equation}
Comparing (\ref{plk}) and (\ref{ckt}) we notice that the densities of
the two shortest paths satisfy $p_{0,k}=kc_k$ and
\hbox{$p_{1,k}=2(k-1)c_k$}. The latter reflects that there are $k-1$
links in a tree of size $k$ and that with unit probability all
components are trees (as discussed in the next section).

Note also that the longest possible path, $l=k-1$, corresponds to
linear (chain-like) components. According to Eq.~(\ref{plk}), the
density of such paths is $p_{k-1,k}=t^{k-1}e^{-kt}$. This density
decays exponentially with length, so these components are typically
small, their length being of the order one.

The path length density can be simplified in the large $k$-limit by
considering the properly normalized ratio of factorials
\begin{eqnarray*}
\frac{k!}{k^l\,(k-l)!}
&=&\prod_{j=1}^{l-1} \left(1-\frac{j}{k}\right)\\
&=& \exp\left(-\sum_{j=1}^{l} \frac{j}{k}
+\frac{1}{2}\sum_{j=1}^{l} \frac{j^2}{k^2}-\ldots\right)\\
&\simeq& \exp(-l^2/2k)\,.
\end{eqnarray*}
Using the Stirling formula, in the limits $k\gg 1$ and $l\gg 1$, the
path density becomes
\begin{equation}
\label{plk-largek}
p_{l,k}\simeq l\,(2\pi k^3)^{-1/2}\,t^{k-1}\,e^{k(1-t)}\,e^{-l^2/2k}.
\end{equation}
As was the case for the component size distribution, the path
length density is self-similar in the vicinity of the gelation
point, $p_{l,k}\to (1-t)^2 \Phi_p\left(k(1-t)^2,l(1-t)\right)$,
with the scaling function
\begin{equation}
\label{phip}
\Phi_p(\xi,\eta)=\eta\,(2\pi \xi^3)^{-1/2}\exp(-\eta^2/2\xi).
\end{equation}
Thus, the characteristic path length diverges near the gelation point, $l\sim
(1-t)^{-1}$.

At the critical point, the path length density becomes
\begin{equation}
\label{plkg}
{\bf p}_{l,k}\simeq l(2\pi k^3)^{-1/2}\,\exp(-l^2/2k).
\end{equation}
It is evident that the typical path length scales as square root of
the component size
\begin{equation}
\label{lk}
l\sim k^{1/2}.
\end{equation}

For finite systems, the scaling law for the typical path length
(\ref{lk}) combined with the characteristic component size (\ref{kg}) leads
to the following characteristic path length
\begin{equation}
\label{ln}
l\sim N^{1/3}.
\end{equation}
One can deduce several other scaling laws and finite-size scaling
functions underlying the path density. For example, substituting the
gelation time $1-t_g\sim N^{-1/3}$ into the total number of paths
$g=(1-t)^{-1}$ yields $g\sim N^{1/3}$.

\section{Cycles}

Each component has a certain number of nodes and links. The complexity
of a component is defined as the number of links minus the number of
nodes.  Components with complexity $-1$ are trees; components with
complexity $0$ and $1$ are termed unicyclic and bicyclic
correspondingly.  Finite components are predominantly trees.  We have
seen that the overall number of links is proportional to $N$ and that
the overall the number of self-links is of the order unity. The
overall numbers of trees and of unicyclic components mirror this
behavior. Generally, the number of components of complexity $R$ is
proportional to $N^{-R}$ (this result is well-known, see
e.g. \cite{jklp,bk1} and especially \cite{alon}).  Therefore, it
suffices to characterize trees and unicyclic components only.

Each unicyclic component contains a single cycle. Cycles are an
important characteristic of a graph \cite{rm,rkbb}.  In this section,
we analyze cycles and unicyclic components using the rate equation
approach. We first note that cycles in random graphs were also studied
using various other approaches: Janson \cite{sj,sj1} employs
probabilistic and combinatorial techniques; Marinari and Monasson
\cite{rm} assign an Ising spin to each node and deduce certain
properties of loops from the partition function of the Ising model;
Burda {\em et al} \cite{bjk} modify a random graph model to favor the
creation of short cycles, and examine the model using a diagrammatic
technique.  A number of authors also studied cycles on information
networks like the Internet (see \cite{internet} and references
therein).

\subsection{Infinite System}

There is a significant difference between the distribution of trees
and unicyclic components. In the thermodynamic limit, the number of
trees is extensive and as a result, it is a deterministic, or a
self-averaging quantity. The number of unicyclic components is not
extensive, but rather of the order unity; as a result it is a random
quantity with a nontrivial distribution even for infinite random
graphs. In what follows, we study the {\it average} number of
unicyclic components of a given size or cycle length.

The average number of cycles follows directly from the path length
density. Quite simply, when the two extremal nodes in a path are
linked, a cycle is born. Let the number of cycles of size $l$ at time
$t$ be $w_l(t)$. It grows according to the rate equation
\begin{equation}
\label{wl-eq}
\frac{dw_l}{dt}=\frac{1}{2}\,q_{l-1}.
\end{equation}
The right-hand side equals the link creation rate $1/(2N)$ times the
total number of paths $Nq_{l-1}$; indeed, the total number of cycles
of a given length is of the order one.  The cycle length distribution
is
\begin{equation}
\label{wl}
w_l=\frac{t^l}{2l}.
\end{equation}
In particular, at the gelation point, the cycle length distribution is
inversely proportional to the cycle length \cite{jklp}
\begin{equation}
\label{wlg}
{\bf w}_l=(2l)^{-1}.
\end{equation}
This result can alternatively be obtained using combinatorics.

To characterize cycles in a given component size, we consider the
joint distribution $u_{l,k}$, the average number of unicyclic
components of size $k$ containing a cycle of length $l$ with $1\leq
l\leq k$. This joint distribution evolves according to the {\it
linear} rate equation
\begin{equation}
\label{ulk-re}
\frac{du_{l,k}}{dt}=\frac{1}{2}p_{l-1,k}+
\sum_{i+j=k}(iu_{l,i})\,(jc_j)-k\,u_{l,k}
\end{equation}
for $l\geq 1$. Initially there are no cycles, and therefore $u_{l,k}(0)=0$.
Eliminating the time dependence via the substitution
\hbox{$u_{l,k}=U_{l,k}t^ke^{-kt}$}, the coefficients satisfy the recursion
\begin{equation}
\label{ulk}
k\,U_{l,k}=\frac{1}{2}\,P_{l-1,k}+\sum_{i+j=k}(iU_{l,i})\,(jC_j).
\end{equation}
Using the generating function $U_l(z)=\sum_k e^{kz} U_{l,k}$ this
recursion is recast into the differential equation
\hbox{$(1-G)\,\frac{dU_l}{dz}=\frac{1}{2}\,P_{l-1}$}. Dividing  by
(\ref{g-eq}), we obtain
\begin{equation}
\label{Ul-eq}
\frac{dU_l}{dG}=\frac{1}{2}G^{l-1}.
\end{equation}
Integrating this equation yields the generating function
\begin{equation}
\label{Ul}
U_l(z)=\frac{1}{2l}\,G^l(z).
\end{equation}
Consequently, the cycle length distribution (in finite components
only) is $p_l=\frac{\tau^l}{2l}$, in agreement with (\ref{wl}) prior
to the gelation time ($t<1$).

Additionally, the joint generating function defined as
$u(z,w)=\sum_{l,k}e^{kz}\,w^l\, u_{l,k}$ is given by
\begin{equation}
\label{uzw}
u(z,w)=\frac{1}{2}\ln \frac{1}{1-wG(z+\ln t-t)}.
\end{equation}
As for paths, statistics of cycles are directly coupled to statistics
of components via the generating function $G(z)$.  The total number of
unicyclic components of finite-size $h=\sum_{l,k} u_{l,k}$ is
therefore
\begin{equation}
\label{utot}
h(t)=\frac{1}{2}\,\ln \frac{1}{1-\tau}.
\end{equation}
Below the gelation point, $h(t)=\frac{1}{2}\ln \frac{1}{1-t}$, for
$t<1$. The total number of unicyclic components can alternatively be
obtained by noting that (i) it satisfies the rate equation
$dh/dt=\frac{1}{2}\sum_k k^2 c_k=\frac{1}{2}\,M_2$, and (ii) the
second moment of the size distribution is $M_2=(1-t)^{-1}$ for $t<1$
as follows from (\ref{czt}).

The coefficients underlying the cycle distribution are found using
contour integration. Indeed, writing \hbox{$U_{l,k}=(2\pi i)^{-1}\oint
U_l\,y^{-k-1}\,dy$} and substituting $r=l$ in (\ref{ak}) gives
$U_{l,k}=\frac{1}{2}\,\frac{k^{k-l-1}}{(k-l)!}$ \cite{jlr}. The cycle
length-size distribution is therefore
\begin{equation}
\label{ulkt}
u_{l,k}(t)=\frac{1}{2}\,\frac{k^{k-l-1}}{(k-l)!}t^ke^{-kt}.
\end{equation}
The smallest cycle, $l=1$, is a self-connection, and the average
number of such cycles is $u_{1,k}=\frac{t}{2}\,kc_k$. The largest
cycles are rings, $l=k$, and their total number is on average
$u_{k,k}=\frac{1}{2k}\,t^k\,e^{-kt}$.

The large-$k$ behavior of the cycle length distribution is found
following the same steps leading to (\ref{plk-largek})
\begin{equation}
\label{ulkt-largek}
u_{l,k}(t)\simeq (8\pi k^3)^{-1/2}\,t^k\,e^{k(1-t)}\,e^{-l^2/2k}.
\end{equation}
This distribution is self-similar in the vicinity of the gelation
transition, \hbox{$u_{l,k}(t)\to
(1-t)^3\Phi_u\left(k(1-t)^2,l(1-t)\right)$}, with the scaling function
\begin{equation}
\label{phiu}
\Phi_u(\xi,\eta)=(8\pi\xi^3)^{-1/2}\exp(-\eta^2/2\xi).
\end{equation}
We see that the cycle length is characterized by the same scale as the
path length, $l\sim (1-t)^{-1}$.  At the gelation point, the
distribution is
\begin{equation}
\label{ulk-tg}
{\bf u}_{l,k}\simeq (8\pi k^3)^{-1/2}\exp(-l^2/2k).
\end{equation}
Fixing the component size, the typical cycle length behaves as the
typical path length, $l\sim k^{1/2}$.

The size distribution of unicyclic components is found from the joint
distribution $v_k=\sum_l u_{l,k}$. Using (\ref{ulkt}) we get \cite{bk1}
\begin{equation}
\label{vk}
v_k(t)=\frac{1}{2}\left(\sum_{n=0}^{k-1}\frac{k^{n-1}}{n!}\right)
t^k\,e^{-kt}.
\end{equation}
This distribution can alternatively be derived from the {\it linear}
rate equation
\begin{equation}
\label{vk-eq}
\frac{dv_k}{dt}=\frac{1}{2}k^2c_k+\sum_{i+j=k}(iv_i)(jc_j)-k\,v_k.
\end{equation}
This equation is obtained from (\ref{ulk-re}) using the equality
$k^2c_k=\sum_l p_{l,k}$. It reflects that linking a pair of nodes
in a component generates a unicyclic component.  Integrating
(\ref{ulk-tg}) over the cycle length, the critical size
distribution of unicyclic components has an algebraic tail
\begin{equation}
\label{qk-tg}
{\bf v}_k\simeq (4k)^{-1}.
\end{equation}

\subsection{Finite Systems}

\begin{figure}[ht]
\includegraphics[width=8cm]{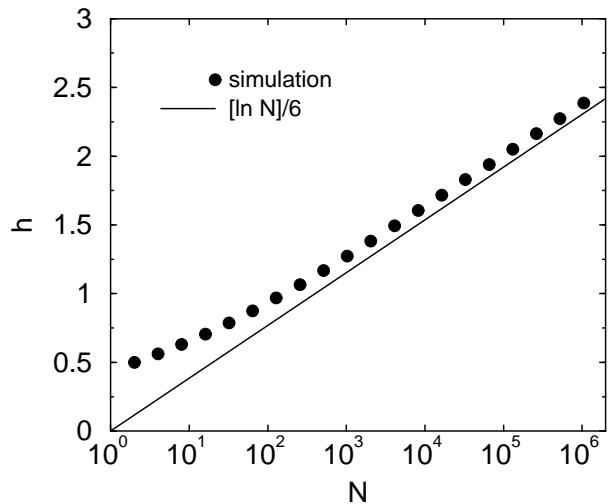}
\caption{The total number of unicyclic components versus the
system size at the gelation point. Shown is $h$ versus $N$. Each
data point represents an average over $10^6$ independent
realizations.} \label{fig-utot}
\end{figure}

We turn now to finite systems, restricting our attention to the
gelation point. The total number of unicyclic components is obtained
by estimating $h(N,t_g)$. Substituting (\ref{tg}) into (\ref{utot})
shows that the average number of unicyclic components (and hence,
cycles) grows logarithmically with the system size
(Fig.~\ref{fig-utot})
\begin{equation}
\label{utotn}
h(N)\simeq \frac{1}{6}\ln N.
\end{equation}

\begin{figure}[ht]
\includegraphics[width=8cm]{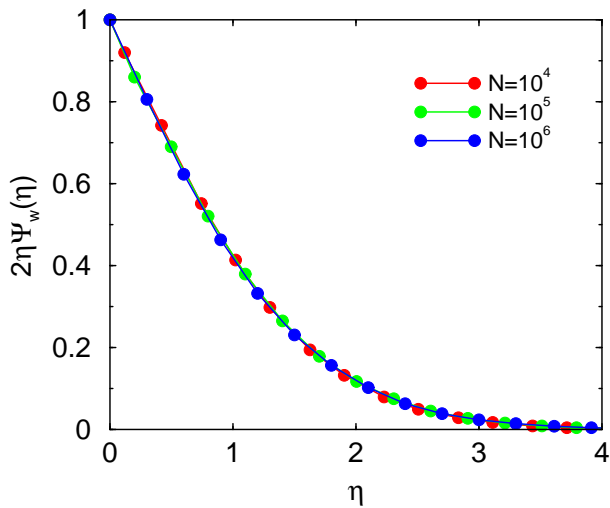}
\caption{Finite-size scaling of the cycle-length distribution.  Shown
is $2\eta\Psi_w(\eta)$ versus $\eta$ obtained using systems with size
$N=10^4$, $10^5$, and $10^6$. The data represents an average over
$10^6$ independent realizations.} \label{fig-pnfs}
\end{figure}

Comparing the path length distribution (\ref{plkg}) and the cycle
length distribution (\ref{ulk-tg}), we conclude that the characteristic cycle
length and the characteristic path length obey the same scaling law, $l\sim
N^{1/3}$. This implies that the cycle length distribution in a finite
system of size $N$, $w_l(N)$, obeys the finite-size scaling law
\begin{equation}
\label{psiw}
{\bf w}_l(N)\sim N^{-1/3}\Psi_w\left(lN^{-1/3}\right).
\end{equation}
Numerical simulations confirm this behavior (Fig.~\ref{fig-pnfs}).

In the simulations, analysis of cycle statistics requires us to keep
track of all links.  Cycles are conveniently identified using the
standard ``shaving'' algorithm. Dangling links, i.e., links involving
a single-link node are removed from the system sequentially. The link
removal procedure is carried until no dangling links remain. At this
stage, the system contains no trees. Simple cycles are those
components with an equal number of links and nodes.

\begin{figure}
\includegraphics[width=8cm]{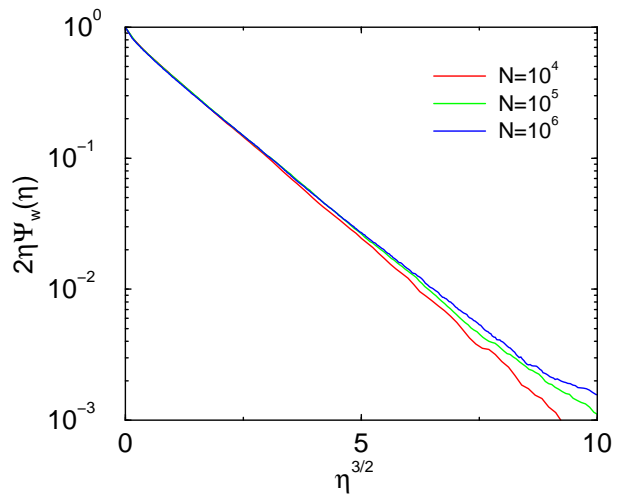}
\caption{The tail of the scaling function.  Shown is
  $2\eta\Psi_w(\eta)$ versus $\eta^{3/2}$.}
\label{fig-pnfs-tail}
\end{figure}

The extremal behaviors of the finite-size scaling function are as follows
\begin{equation}
\label{psiw-ext}
\Psi_w(\eta)\simeq
\begin{cases}
(2\eta)^{-1}&\eta\to 0,\\
\exp(-C\,\eta^{3/2})&\eta\to\infty.
\end{cases}
\end{equation}
The small-$\eta$ behavior follows from (\ref{wlg}).  Statistics of
extremely large cycles can be understood by considering the largest
possible cycles. When there are $n=N/2$ links, the largest possible
cycle has length $l=N/2$. Its likelihood $w(n,2n)$ is obtained using
combinatorics
\begin{equation}
\label{wn2n}
w(n,2n)={2n\choose n}\times \frac{n!}{2n}\times (2n)^{-n}.
\end{equation}
There are ${2n\choose n}$ ways to choose the nodes participating in
the cycle and the next term is the number of ways to arrange them in a
cycle. The corrective factor $2n$ accounts for rotation and reflection
symmetries. The last term is the probability that each pair of
consecutive nodes are linked. The large-$n$ asymptotic behavior is
\begin{equation}
\label{wn2n-largen}
w(n,2n)\simeq \frac{1}{\sqrt{2}n}\left(\frac{2}{e}\right)^n.
\end{equation}
Therefore, $w(n,2n)\sim \exp(-C\, N)$. Substituting $l\sim N$ into the
scaling form (\ref{psiw}) leads to the super-exponential behavior
$\Psi_w(\eta)\sim \exp(-C\eta^{3/2})$, see Fig.~\ref{fig-pnfs-tail}.

\begin{figure}[ht]
\includegraphics[width=8cm]{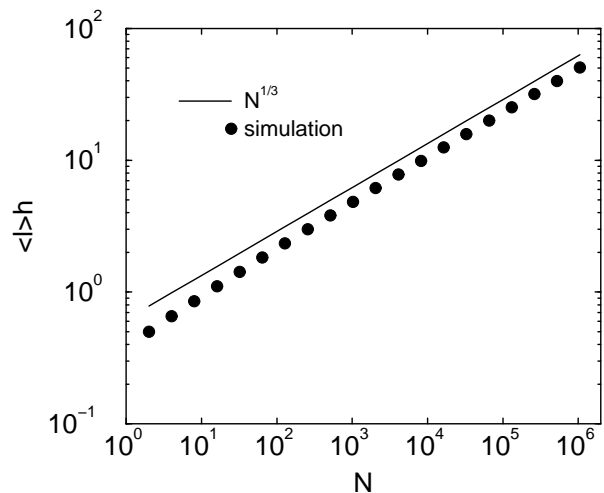}
\caption{The average cycle size at the gelation point. Shown is
$\langle l(N)\rangle h(N)$ versus $N$.  Each data point
represents an average over $10^6$ independent realizations.}
\label{fig-lav}
\end{figure}

Typically, cycles are of size $N^{1/3}$. The average moments $\langle
 l(N)\rangle=\sum_l\,l\, w_l(N)/\sum_l w_l(N)$ reflect this
 law. However, the algebraic divergence, $w_l\sim l^{-1}$, leads to a
 logarithmic correction as follows from
 (\ref{utotn})--(\ref{psiw-ext}):
\begin{equation}
\langle l^n(N)\rangle \sim N^{n/3}[\ln N]^{-1}.
\end{equation}
The behavior of the average cycle length is verified numerically
(Fig.~\ref{fig-lav}).

Finite-size scaling of other cycle statistics such as the joint
distribution can be constructed following the same procedure. For
example, the size distribution of unicyclic components should follow
the scaling form
\begin{equation}
v_k(N) \sim N^{-2/3}\Psi_v\left(kN^{-2/3}\right).
\end{equation}
The scaling function diverges $\Psi_v(\xi)\simeq (4\xi)^{-1}$ for
$\xi\to 0$.

\section{The first cycle}

The above statistical analysis of cycles characterizes the average
behavior but not necessarily the typical one because the number of
cycles is a fluctuating quantity. There are numerous interesting
features concerning cycles that are not captured by the average number
of cycles.  For instance, what is the probability that the system does
not contain a cycle up to time $t$?  It suffices to answer this
question in the pre-gel regime as the giant component certainly
contains cycles.

\begin{figure}[ht]
\includegraphics[width=8cm]{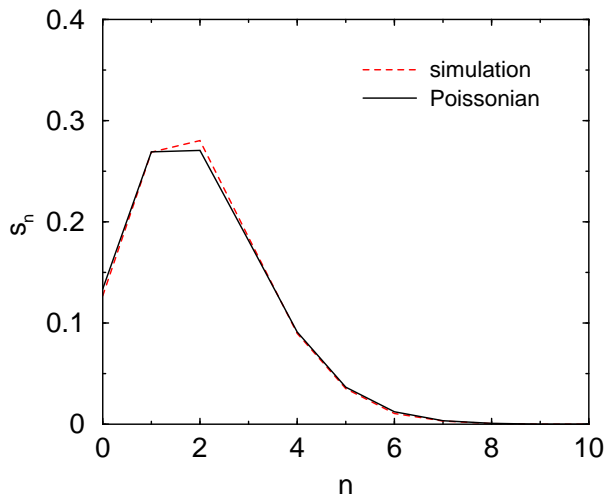}
\caption{The distribution of the number of cycles.  Shown is $s_n$
  versus $n$ at the gelation point. The system size is $N=10^5$ and an
  average over $10^5$ realizations has been performed. A Poissonian
  distribution with an identical average is also shown for reference.}
\label{fig-poisson}
\end{figure}

Let $s_0(t)$ be the (survival) probability that the system does not
contain a cycle at time $t$. The cycle production rate is
$J=\frac{dh}{dt}=\frac{1}{2(1-t)}$.  The number of cycles is finite in
the pre-gel regime, since cycles are independent of each other in the
$N\to\infty$ limit. This assertion (supported by numerical
simulations, see Fig.~\ref{fig-poisson}) implies that the cycle
production process is completely random. The cycle production rate
characterizes the survival probability $s_0$ as follows
\begin{equation}
\label{s0-eq}
\frac{ds_0}{dt}=-Js_0.
\end{equation}
The initial condition is $s_0(0)=1$. As a result, the survival
probability is
\begin{equation}
\label{s0}
s_0(t)=(1-t)^{1/2}
\end{equation}
for $t\leq 1$.  The survival probability vanishes beyond the gelation
point, $s_0(t)=0$ for $t>1$. This reiterates that in the thermodynamic
limit, a cycle is certain to form prior to the gelation transition
\cite{jklp}.

Since the number of cycles produced is of the order of one in the
pre-gel regime, one may expect that statistical properties of cycles
strongly depend on their generation number or alternatively on their
creation time. This is manifested by the first cycle. The quantity
$dt\,s_0\,\frac{dw_l}{dt}$ is the probability that (i) the system
contains no cycles at time $t$, (ii) a cycles is produced during the
time interval $(t,t+dt)$, and (iii) its length is $l$. Summing these
probabilities gives the probability that the first cycle produced
sometimes during the pre-gel regime has length $l$:
\begin{equation}
\label{fl-int}
f_l=\int_0^1 dt\, s_0\, \frac{dw_l}{dt}=\frac{1}{2}\int_0^1 dt\,
(1-t)^{1/2}\,t^{l-1}.
\end{equation}
Summing these quantities, we verify the normalization
\begin{equation*}
\sum_{l\geq 1}f_l=\frac{1}{2}\int_0^1 dt\,(1-t)^{-1/2}=1.
\end{equation*}
The length distribution of the first cycle can be expressed in
terms of the beta function $f_l=\frac{1}{2}\,B(3/2,l)$ or
alternatively
\begin{equation}
\label{fl}
f_l=\frac{\sqrt{\pi}}{4}\,\frac{\Gamma(l)}{\Gamma(l+3/2)}.
\end{equation}
The probability distribution $f_l$ has an algebraic tail,
\begin{equation}
\label{fl-tail}
f_l\simeq C\,l^{-3/2}, 
\end{equation}
with $C=\frac{\sqrt{\pi}}{4}$ for $l\gg 1$. The tail exponent
characterizing the distribution of the first cycle is larger compared
with the exponent characterizing all cycles, reflecting the fact that
the first cycle is created earlier.

Similarly, one can obtain additional properties of the first
cycle. We mention the probability $F_k$ that the first unicyclic
component has size $k$,
\begin{equation}
\label{Fk-int}
F_k=\int_0^1 dt\, s_0\, \frac{1}{2}\,k^2c_k=
\frac{1}{2}\,\frac{k^k}{k!}\,I_k
\end{equation}
with the integral \hbox{$I_k=\int_0^1 dt
(1-t)^{1/2}t^{k-1}e^{-kt}$}. This integral can be expressed in terms
of the confluent hypergeometric function. Its asymptotic behavior can
be readily found by noting that the integrand has a sharp maximum in
the region $1-t\sim k^{-1/2}$ leading to \hbox{$I_k\simeq
2^{-1/4}\Gamma(3/4)k^{-3/4}e^{-k}$}.  Using this in conjunction with
the Stirling's formula, the size distribution has the algebraic tail
\begin{equation}
\label{Fk-tail}
F_k\simeq C\,k^{-5/4}
\end{equation}
with $C=2^{-7/4}\pi^{-1/2}\Gamma(3/4)$ for $k\gg 1$.

\begin{figure}[ht]
\includegraphics[width=8cm]{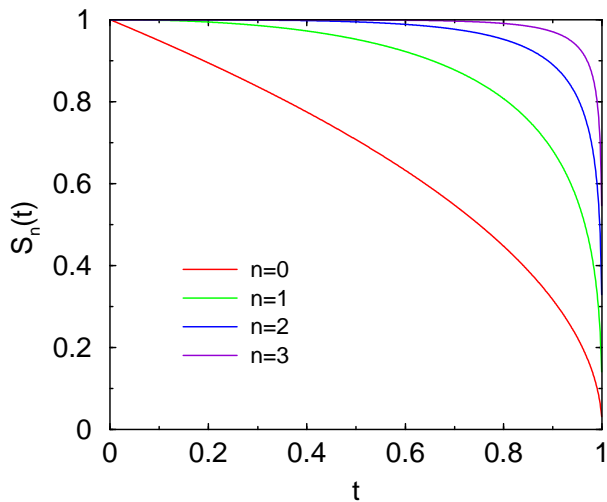}
\caption{The cumulative distribution $S_n(t)=\sum_{0\leq j\leq
n}s_j(t)$ versus $t$ for $n=0,1,2,3$.} 
\label{fig-sn}
\end{figure}

Under the assumption that cycle production is completely random, the number
of cycles obeys Poisson statistics. The probability that there are $n$
cycles, $s_n$, then satisfies the straightforward generalization of
Eq.~(\ref{s0-eq}), viz. $\frac{ds_n}{dt}=J[s_{n-1}-s_n]$ with the initial
condition $s_n(0)=\delta_{n,0}$. The solution is the Poisson distribution
$s_n=\frac{h^n}{n!}e^{-h}$, see Fig.~\ref{fig-poisson}.  Explicitly, the
distribution reads
\begin{equation}
\label{sn}
s_n=\frac{(1-t)^{1/2}}{n!}\left[\frac{1}{2}\ln \frac{1}{1-t}\right]^n.
\end{equation}
The cumulative distribution $S_n(t)=s_0(t)+\ldots+s_n(t)$ is plotted
 in Fig.~\ref{fig-sn}.

The Poisson distribution (\ref{sn}) can also be used to calculate $f_{n,l}$
the size distribution of the $n$th cycle. We merely quote the large-$l$ tail
behavior
\begin{equation}
\label{fnl}
f_{n,l}\sim \frac{1}{(n-1)!}\,l^{-3/2}\left[\frac{1}{2}\ln l\right]^{n-1}
\end{equation}
Indeed, summation over the cycle generation reproduces the overall
cycle distribution (\ref{wlg}).

\begin{figure}[ht]
\includegraphics[width=8cm]{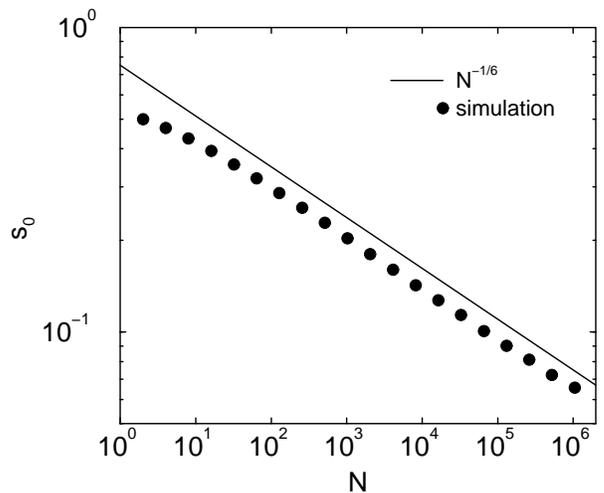}
\caption{The survival probability versus the system size. Shown is
$s_0(N)$ versus $N$ at the gelation point, i.e., when $N/2$ links are
placed. Each data point represents an average over $10^6$
realizations.}
\label{fig-s0}
\end{figure}

In finite systems, it is possible that no cycle are created at the
gelation time. This probability decreases algebraically with the
system size, as seen by substituting (\ref{tg}) into (\ref{s0})
\begin{equation}
\label{s0n}
s_0\sim N^{-1/6}.
\end{equation}
This prediction agrees with simulations, see Fig.~\ref{fig-s0}.  In
practice, this slow decay indicates that a relatively large system may
contain no cycles after $N/2$ links are placed. Generally, the
probability that there is a finite number of cycles increases with the
number of cycles
\begin{equation}
\label{snn}
s_n\sim \frac{1}{n!}\,N^{-1/6}\left[\frac{1}{6}\ln N\right]^n.
\end{equation}

The length distribution of the first cycle is characterized by the
same $l\sim N^{1/3}$ size scale as does the overall cycle
distribution. We focus on the behavior of the moments
\begin{equation}
\label{ln1}
\langle l^n\rangle \sim N^{n/3-1/6}.
\end{equation}
This behavior is obtained from the distribution (\ref{fl-tail}) that
should be integrated up to the appropriate cutoff, i.e., $\langle
l^n\rangle \sim \int_1^{N^{1/3}} dl\, l^n\, l^{-3/2}$. As a result,
the average size of the first cycle is much smaller than the characteristic
cycle size $\langle l\rangle \sim N^{1/6}$.  Moments corresponding to
the size of the first unicyclic component grow as follows
\begin{equation}
\label{kn1}
\langle k^n\rangle \sim N^{2n/3-1/6},
\end{equation}
as obtained from (\ref{Fk-tail}). Consequently, the average size
of the first unicyclic component is smaller than the characteristic
component size, $\langle k\rangle \sim N^{1/2}$.

\section{Conclusions}

In summary, we have extended the kinetic theory description of random
graphs to structures such as paths and cycles. Modeling the linking
process dynamically leads to an aggregation process for both
components and paths. The density of paths in finite components is
coupled to the component size distribution via nonlinear rate
equations while the average number of cycles is coupled to the path
density via linear rate equations. Both path and cycle length
distributions are coupled to the component size distribution.

Generally, size distributions decay exponentially away from the
gelation point, but at the gelation time, algebraic tails emerge. As
the system approaches this critical point, the size distributions
follow a self-similar behavior characterized by diverging size scales.

The kinetic theory approach is well-suited for treating infinite
systems. The complementary behavior for finite systems can be obtained
from heuristic scaling arguments. This approach yields scaling laws
for the typical component size, path length, and cycle length at the
gelation point. These scaling laws can be formalized using finite-size
scaling forms, i.e., self-similarity as a function of the system size,
rather than time. Obtaining the exact form of these scaling functions
is a nice challenge in particular for the most fundamental quantity,
the component size distribution that is characterized by a
non-monotonic scaling function.

The kinetic theory approach seems artificial at first sight. Indeed,
graphs are discrete in nature and therefore combinatorial approaches
appear more natural. Yet, once the rate equations are formulated, the
analysis is straightforward. Utilizing the continuous time variable
allows us to employ powerful analysis tools. Moreover, some of the
kinetic theory results are less cumbersome compared with the
combinatorial results.

The same methodology can be expanded to analyze other features of
random graphs. For example, correlations between the node degree and
the cluster size can be analyzed using bi-aggregation rate equations
\cite{bk2}. It is quite possible that structural properties in other
aggregation processes, for example, polymerization with a sum kernel
\cite{fl}, and in other variants of random graphs such as small-world
networks \cite{ws} can be analyzed using kinetic theory.

One could try to utilize kinetic theory to probe the distribution of
various families of subgraphs.  We have limited ourselves to cycles
since they, alongside with trees, do appear in random graphs while
more interconnected families of subgraphs are very rear
\cite{alon}. Yet in biological and technological networks certain
interconnected families of subgraphs do appear. Such populated
families of subgraphs, motifs, are believed to carry information
processing functions \cite{alon1,victor}. It will be interesting to
use kinetic theory to analyze motifs in special random graphs.

\bigskip

This research was supported by the DOE (W-7405-ENG-36).

\appendix

\section{Contour integration}
Let $A(z)=\sum_k A_k e^{kz}$ be the generating function of the
coefficients $A_k$. For the family of generating functions 
$A(z)=G^r(z)$ with $G(z)$ satisfying $Ge^{-G}=e^z$, the coefficients
$A_k$ can be obtained via contour integration in the complex $y$ plane
where $y=e^z$ as follows
\begin{eqnarray}
\label{ak}
A_k&=&\frac{1}{2\pi i}\oint dy \,\frac{G^r}{y^{k+1}}\nonumber\\
       &=&\frac{1}{2\pi i}\oint dG \,G^r\,
          \frac{e^{(k+1)G}}{G^{k+1}}\frac{dy}{dG}\nonumber\\
       &=&\frac{1}{2\pi i}\oint dG \,G^r\,
          \frac{e^{(k+r)G}}{G^{k+1}}(1-G)e^{-G}\nonumber\\
       &=&\frac{1}{2\pi i}\oint dG
          \sum_n \frac{k^n}{n!}(G^{n+r-k}-G^{n+r+1-k})\nonumber\\
       &=&r\frac{k^{k-r-1}}{(k-r)!}.
\end{eqnarray}
Since $Ge^{-G}=e^z$, it is convenient to perform the integration in
the complex $G$ plane. In writing the third line, we used
$\frac{dy}{dG}=(1-G)e^{-G}$.


\begin{thebibliography}{99}

\bibitem{sr}
   R.~Solomonoff and A.~Rapaport,
   Bull. Math. Biophys. {\bf 13}, 107 (1959).

\bibitem{er}
   P.~Erd\H os and A.~R\'enyi,
   Publ.\ Math.\ Inst.\ Hungar.\ Acad.\ Sci. {\bf 5}, 17 (1960).

\bibitem{bb}
   B.~Bollob\'as,
   {\em Random Graphs} (Academic Press, London, 1985).

\bibitem{jlr}
   S.~Janson, T.~\L uczak, and A.~Rucinski,
   {\em Random Graphs} (John Wiley \& Sons, New York, 2000).

\bibitem{jklp}
   S.~Janson, D.~E.~Knuth, T.~\L uczak, and B.~Pittel,
   Rand.\ Struct.\ Alg. {\bf 3}, 233 (1993).

\bibitem{pjf}
   P.~J.~Flory, J. Amer. Chem. Soc. {\bf 63}, 3083 (1941).

\bibitem{whs}
   W.~H.~Stockmayer, J. Chem. Phys. {\bf 11}, 45 (1943).

\bibitem{pjf1}
   P.~J.~Flory, {\em Principles of Polymer Chemistry}
   (Cornell University Press, Ithaca, 1953).

\bibitem{ds}
   D.~Stauffer, {\em Introduction to Percolation Theory}
   (Taylor \& Francis, London, 1985).

\bibitem{kcbh}
   T.~Kalisky, R.~Cohen, D.~ben-Avraham, and S.~Havlin, 
   Lect. Notes. Phys. {\bf 650}, 3 (2004).

\bibitem{bbckw}
   B.~Bollob\'as, C.~Borgs, J.~T.~Chayes, J.~H.~Kim, and D.~B.~Wilson,
   Rand.\ Struct.\ Alg. {\bf 18}, 201 (2001).

\bibitem{nsw}
   M.~E.~J.~Newman, S.~Strogatz, and D.~J.~Watts,
   Phys. Rev. E {\bf 64}, 026118 (2001).

\bibitem{gn}
   M.~Girvan and M.~E.~J.~Newman,
   Proc.\ Natl.\ Acad.\ Sci.\ {\bf 99}, 7821 (2002).

\bibitem{mvs}
   M.~V.~Smoluchowski, Physik. Zeits. {\bf 17}, 557 (1916). 
   Zeits. Phys. Chem. {\bf 92}, 129 (1917).

\bibitem{chandra}
   S.~Chandrasekhar,  Rev.\ Mod.\ Phys. {\bf 15}, 1--89 (1943).

\bibitem{da}
   D.~J.~Aldous,
   Bernoulli {\bf 5}, 3 (1999).

\bibitem{fl}
   F.~Leyvraz,
   Phys. Rep. {\bf 383}, 95 (2003).

\bibitem{jbm}
   J.~B.~McLeod,
   Quart.\ J. Math.\ Oxford {\bf 13}, 119 (1962);
   {\it ibid} {\bf 13}, 193 (1962);
   {\it ibid} {\bf 13}, 283 (1962).

\bibitem{hez}
   E.~M.~Hendriks, M.~H.~Ernst, and R.~M.~Ziff,
   J. Stat.\ Phys.\ {\bf 31}, 519 (1983).

\bibitem{bk}
   E.~Ben-Naim and P.~L.~Krapivsky,
   Europhys.\ Lett. {\bf 65}, 151 (2004).

\bibitem{bk1}
   E.~Ben-Naim and P.~L.~Krapivsky,
   J. Phys.\ A {\bf 37}, L189 (2004).

\bibitem{aal}
   A.~A.~Lushnikov,
   J.\ Colloid.\ Inter.\ Sci.\ {\bf 65}, 276 (1977).

\bibitem{ve}
   P.~G.~J.~van Dongen and M.~H.~Ernst,
   J. Stat.\ Phys.\ {\bf 49}, 879 (1987).

\bibitem{zm}
   D.~H.~Zanette and S.~C.~Manrubia, Physica A {\bf 295}, 1 (2001).

\bibitem{dms}
   S.~N.~Dorogovtsev J.~F.~F.~Mendes, and A.~N.~Samukhin,
   Phys.\ Rev.\  E {\bf 63}, 062101 (2001).

\bibitem{bck}
   Z.~Burda, J.~D.~Correia, and A.~Krzywicki,
   Phys.\ Rev.\ E {\bf 64}, 046118 (2001).

\bibitem{kr}
   P.~L.~Krapivsky and S.~Redner, J. Phys.\ A {\bf 35}, 9517 (2003).

\bibitem{kb}
   P.~L.~Krapivsky and E.~Ben-Naim,
   Phys.\ Rev.\ E {\bf 53}, 291 (1996).

\bibitem{alon}
   S.~Itzkovitz, R.~Milo, N.~Kashtan, G.~Ziv, and U.~Alon,
   Phys.\ Rev.\ E {\bf 68}, 026127 (2003).

\bibitem{rkbb}
   H.~D.~Rozenfeld, J.~E.~Kirk, E.~M.~Bollt, and D.~ben-Avraham,
   {\it cond-mat/0403536}.

\bibitem{rm}
   E.~Marinari and R.~Monasson,
   {\it cond-mat/0407253}.

\bibitem{sj}
   S.~Janson, Rand.\ Struc.\ Alg. {\bf 17}, 343 (2000).

\bibitem{sj1}
   S.~Janson, Combin.\ Probab.\ Comput. {\bf 12}, 27 (2003).

\bibitem{bjk}
   Z.~Burda, J.~Jurkiewicz, and A.~Krzywicki,
   Phys.\ Rev.\ E {\bf 69}, 026106 (2004); {\it ibid}
   {\bf 70}, 026106 (2004).

\bibitem{internet}
   G.~Bianconi, G.~Caldarelli, and A.~Capocci,
   {\it cond-mat/0408349}.

\bibitem{bk2}
   E.~Ben-Naim and P.~L.~Krapivsky,
   unpublished.

\bibitem{ws}
   D.~J.~Watts and S.~H.~Strogatz,
   Nature {\bf 393}, 440 (1998).

\bibitem{alon1} 
    R.~Milo {\em et al}, 
    Science {\bf 298}, 824--827 (2002); 
    {\it ibid} {\bf 303}, 1538--1542 (2004).

\bibitem{victor}
   V.~Spirin and L.~A.~Mirny,
   Proc.\ Natl.\ Acad.\ Sci. {\bf 100}, 12123--12128 (2003).

\end{thebibliography}
\end{document}